\documentclass{article}
\usepackage[utf8]{inputenc}
\usepackage{amsmath}
\usepackage{amssymb}
\usepackage{mathtools}
\usepackage{color}
\usepackage{multirow}
\makeatletter \@addtoreset{equation}{section} \makeatother

\newcommand{\red}[1]{{\color{red} #1}} 
\newcommand{\blue}[1]{{\color{blue} #1}}
\newcommand{\green}[1]{{\color{green} #1}}

\newcommand{\cf}{c.f.\ }
\newcommand{\eg}{e.g.\ }
\newcommand{\ie}{i.e.\ }

\newcommand{\noi}{\vspace{12pt}\noindent}

\newcommand{\eq}[1]{{eq.\ (\ref{#1})}}

\newcommand{\C}{\mathbb{C}}
\newcommand{\N}{\mathbb{N}}
\newcommand{\R}{\mathbb{R}}
\newcommand{\Z}{\mathbb{Z}}

\newcommand{\Wslash}{W\!\!\!\!\!/ \ }
\newcommand{\beq}{\begin{equation}}
\newcommand{\eeq}{\end{equation}}
\newcommand{\bea}{\begin{eqnarray}}
\newcommand{\eea}{\end{eqnarray}}

\newcommand{\tr}{{\rm tr}}

\renewcommand{\~}{ \ }
\renewcommand{\=}{ \ = \ }

\renewcommand{\Re}{{\rm Re}}
\renewcommand{\Im}{{\rm Im}}
\renewcommand{\span}{{\rm span}}

\newcommand{\proofbox}{\begin{flushright}{\hfill \ensuremath{\Box}}
\end{flushright}}

\newtheorem{theorem}{Theorem}[section]

\newtheorem{definition}[theorem]{Definition}

\begin{document}
\thispagestyle{empty}
\title{\Large{\bf Generalized Jarlskog Invariants, \\ 
Mass Degeneracies and Echelon Crosses}}
\author{{\sc Klaus~Bering}$^{a}$ \\~\\
Institute for Theoretical Physics \& Astrophysics\\
Masaryk University\\Kotl\'a\v{r}sk\'a 2\\CZ--611 37 Brno\\Czech Republic}
\maketitle
\vfill
\begin{abstract}
It is known that the Cabibbo-Kobayashi-Maskawa (CKM) $n\times n$ matrix can 
be represented by a real matrix iff there is no CP-violation, and then the 
Jarlskog invariants vanish. We investigate sufficient conditions for the 
opposite statement to hold, paying particular attention to degenerate cases.
We find that higher Jarlskog invariants are needed for $n\geq 4$. One generic
sufficient condition is provided by the existence of a so-called echelon 
cross. 
\end{abstract}
\vfill
\begin{quote}
PACS number(s): 11.30.Er; 11.30.Hv; 12.15.Ff; 12.60.-i; 
\hrule width 5.cm \vskip 2.mm \noindent 
$^{a}${\small E--mail:~{\tt bering@physics.muni.cz}} \\
\end{quote}

\newpage
\tableofcontents

\section{The Setting}

\subsection{Generalized CKM matrix}

In the standard model, Cabibbo-Kobayashi-Maskawa (CKM) matrices 
\cite{c63,km73} are unitary square 
matrices.\footnote{There is an analogous story for the 
Pontecorvo-Maki-Nakagawa-Sakata (PMNS) matrices \cite{p57,mns62,p68} in the 
leptonic sector.} We will sometimes restrict to this physical case below, but 
it is useful to consider a more general setting. 

\noi
Let $n_u,n_d\in\N$ be the number of up- and down-type quarks, 
respectively. We will consider (possibly rectangular, not necessarily unitary)
CKM $n_u\times n_d$ matrices 
\beq V\=(V^i{}_a)\~\in\~{\rm Mat}_{n_u\times n_d}(\C), \eeq
where $i\in \{1,\ldots, n_u\}$ and $a \in \{1,\ldots, n_d\}$.

\subsection{Quark Masses}

Let the up and down mass matrices be (possibly degenerate) diagonal matrices
\beq \begin{aligned} M_u
\=&{\rm diag}(m^u_i)\~\in\~{\rm Mat}_{n_u\times n_u}(\R), \cr
M_d\=&{\rm diag}(m^d_a)\~\in\~{\rm Mat}_{n_d\times n_d}(\R).
 \end{aligned} \eeq
The pertinent terms in the standard model Lagrangian density are\footnote{See 
\eg eq.\ (29.56) in Ref.~\cite{Schw14}. Recall that 
$\bar{\psi}:=\psi^{\dagger}\gamma_0$ and
$\psi_{R/L}:=\frac{1}{2}(1\pm \gamma_5)\psi$.} 
\beq \Delta {\cal L} 
\= \frac{g}{\sqrt{2}}\bar{u}_i^L  V^i{}_a \Wslash^+ d_L^a 
 - m^u_i \bar{u}^L_i u_R^i - m^d_a \bar{d}^L_a d_R^a  + {\rm h.c.}   \eeq
in the mass basis.
\subsubsection{Disclaimer} 

Note that in physics true degeneracy can usually only happen if it is caused 
by a corresponding symmetry in the theory. In this article, we will not 
speculate about such underlying causes. Rather we will treat degeneracy as 
an idealized well-defined mathematical statement, as opposed to a fuzzy 
experimental fact with corresponding error bars, areas of unitarity triangles, 
etc. The point of view of this paper is that the pure mathematical matrix 
problem is interesting in it own right.

\subsection{Residual Global Flavor Symmetry}

There is a residual global flavor symmetry group\footnote{A general CKM matrix 
$V$ commutes with at least a $U(1)$-factor, cf. Schur's Lemma, so the effective
symmetry group has 1 DOF less.} $G\equiv G_u\times G_d$ defined via the 
commutant/isotropy/stabilizer groups
\beq \begin{aligned} 
G_u\~:=\~&\{U_u\in U(n_u) \mid [ U_u, M_u] =0 \}, \cr  
G_d\~:=\~&\{U_d\in U(n_d) \mid [ U_d, M_d] =0 \}. \end{aligned} \eeq
It acts as\footnote{We restrict to subgroups of unitary (rather than general
linear) groups in order to be able to define generalized Jarlskog invariants
with the help of Hermitian adjoint rather than inverse matrix operation 
(which doesn't exist for rectangular matrices). See also a related discussion
in subsection~\ref{sec2222}.}
\beq \begin{aligned} 
V^{\prime}\=U_uVU^{\dagger}_d
\quad\Leftrightarrow&\quad
V^{\prime i}{}_a \= (U_u)^i{}_j\~V^j{}_b\~ (U^{\dagger}_d)^b{}_a, \cr
u^{\prime}_{L/R} \= U_u u_{L/R}
\quad\Leftrightarrow&\quad
u^{\prime i}_{L/R} \= (U_u)^i{}_j\~ u_{L/R}^j, \cr
d^{\prime}_{L/R} \= U_d d_{L/R} 
\quad\Leftrightarrow&\quad
d^{\prime a}_{L/R} \= (U_d)^a{}_b \~d_{L/R}^b.
\end{aligned} \eeq
If the quark masses are non-degenerate, then $G_u=U(1)^{n_u}$ and 
$G_d=U(1)^{n_d}$. 

\noi
If masses are, say, $r$-fold degenerated, then there will be a corresponding 
$U(r)$-factor enhancement in the symmetry group, and so forth. In general, the 
groups $G_u,G_d$ are products of unitary groups.

\begin{definition}
Define an {\bf equivalence relation} among CKM matrices
\beq V\~\sim\~V^{\prime}
\quad\Leftrightarrow\quad 
\exists U_u\in G_u, U_d\in G_d:\~\~V^{\prime}\=U_uVU^{\dagger}_d,
\eeq
so that CKM matrices are {\bf equivalent} iff they belong to the same $G$-orbit.
\end{definition}

\noi
It turns out that {\em the model has no $CP$-violation iff the CKM matrix $V$ 
is equivalent to a real matrix.}

\noi
Famously, a unitary $3\times 3$ CKM matrix $V$ is equivalent to a real matrix 
iff the Jarlskog invariants vanish, which we revisit in 
subsection~\ref{sec33uni}. In this paper, we investigate possible 
generalizations of this statement.

\subsection{Unitary Matrix Decomposition}
\label{secunitarymatrixdecomp}
\noi
We're in the business of trying to make complex CKM matrices real by acting 
with unitary matrices.
It is often useful to parametrize the unitary matrices as
\beq U(n) \~\ni\~U\=\underbrace{e^{iS}}_{\text{"phases"}} 
\underbrace{O}_{\scriptsize \begin{array}{c}\text{"Euler}\cr\text{angles"} \end{array}}, \eeq
where the matrix $O\in O(n)$ is orthogonal and the matrix $S$ belongs to the 
$\frac{1}{2}n(n\!+\!1)$-dimensional vector space
\beq s(n) \~ := \~ \{ S\in {\rm Mat}_{n\times n}(\R) \mid S^T=S\} \eeq 
of real symmetric matrices. Also note that
\beq \forall O\~ \in \~ O(n):\~ \~ O s(n) O^{-1} \~ \subseteq \~ s(n). \eeq
Since the orthogonal matrices $O$ preserve real matrices, they are not that 
useful to us. The important role is instead played by the real symmetric 
matrices $S$.

\subsection{Double Commutant}

Define the double commutants
\beq \begin{aligned} 
V_u\~ := \~ &\{D_u\in {\rm Mat}_{n_u\times n_u}(\R) \mid 
\forall U_u\in G_u:  [ U_u, D_u] =0 \}, \cr  
V_d\~ := \~ &\{D_d\in {\rm Mat}_{n_d\times n_d}(\R) \mid 
\forall U_d\in G_d:  [ U_d, D_d] =0 \}. 
\end{aligned} \eeq
Note that the double commutants $V_u, V_d$ are finite dimensional real vector 
spaces. The elements $D_u,D_d$ consist of only diagonal matrices. The 
eigenvalues/diagonal elements are degenerate if the corresponding quark masses 
are degenerated. 

\noi
It will be enough to consider bases $(P^u_i)_i$ and $(P^d_a)_a$ of the 
appropriate projection matrices for the double commutants $V_u, V_d$. If the 
mass matrices are of the form\footnote{Here we are slightly misusing the 
notation by not introducing a new summation index in degenerate cases where 
the number of summands are smaller. Hopefully it does not lead to confusion.}
\beq M_u \=\sum_i m^u_i P^u_i, \qquad M_d 
\= \sum_a m^d_a P^d_a, \eeq
then the double commutants are spanned by the corresponding projections
\beq V_u\=\span_{\R}\{P^u_i | i\}, \qquad V_d
\= \span_{\R}\{P^d_a | a\},\eeq
respectively.

\subsection{Jarlskog Invariants}

\begin{definition}
Given a CKM matrix $V$ the {\bf Jarlskog invariants}\footnote{Technically, 
what  is called {\bf Jarlskog invariants} in this paper generalizes the usual 
quartic Jarlskog invariants \cite{j85a,j85b,js88,bn87,j89}. The quadratic 
Jarlskog invariants 
\beq J(I;A) \~ := \~ \Im(\tr(IVAV^{\dagger}))  \~ = \~  0 \eeq 
vanish identically, and are hence not useful. There is a straightforward 
generalization to higher Jarlskog invariants 
\beq\begin{aligned}
J(I,J,K, \ldots,P ;A,B,C,\ldots,H)
 \~ := \~ &\Im(\tr(IVAV^{\dagger} J V B V^{\dagger}K V C V^{\dagger}\ldots 
V^{\dagger}P V H V^{\dagger})) \cr
 \~ = \~ & J(J,K, \ldots,P,I ;B,C,\ldots,H,A) \cr
 \~ = \~ & -J(I,P,\ldots ,K,J;H,\ldots, C,B,A)
,\end{aligned}\eeq 
of even order, which we will only discuss sporadically in footnotes. It is 
natural to speculate that if Jarlskog invariants of all orders vanish then 
the CKM matrix $V$ is equivalent to a real matrix.} are a multi-linear map
\beq \bigwedge \! {}^2V_u \otimes \bigwedge \! {}^2V_d 
 \~ \~ \stackrel{J}{\longrightarrow} \~ \~ \R \eeq
defined as
\beq\begin{aligned} 
J(I,J;A,B) \~ := \~ &\Im(\tr(IVAV^{\dagger} J V B V^{\dagger}))\cr 
 \= &-\Im(\tr(IVAV^{\dagger} J V B V^{\dagger})^{\dagger})\cr 
 \= &-\Im(\tr(VBV^{\dagger} J V A V^{\dagger}I))\cr 
 \= &-\Im(\tr(IVBV^{\dagger} J V A V^{\dagger}))\cr 
 \= &-J(I,J;B,A)\cr
 \= &-J(J,I;A,B),
\end{aligned} \eeq
where $I,J\in V_u$ and $A,B\in V_d$.
\end{definition}
\noi
It is enough to specify the Jarlskog invariants $J(P^u_i,P^u_j;P^d_a,P^d_b)$ on 
a basis of projection matrices for $V_u, V_d$. These are typically labelled by 
via corresponding row and column indices as a shorthand notation. 

\noi
{\em The Jarlskog invariants are $G$-invariant, and they vanish $J=0$ if $V$ 
is equivalent to a real matrix.} The main purpose of this paper is to 
investigate the opposite relationship.

\subsubsection{Unitary Case: Linear Relations}
\label{seclinrel}

Then $J(I,J;A,B)$ vanishes if one of its four entries $I,J,A,B$ is an identity 
matrix. Since the sum of the basis of projection matrices is the identity 
matrix, this leads to identities \cite{js88,cojwk20} among the Jarlskog 
invariants:
\beq\begin{aligned}  
0\=&J\left( {\bf 1}_{n_u\times n_u}, J; A,B\right)
\=\sum_i J\left( P^u_i, J; A,B\right),\cr
0\=&J\left(I, {\bf 1}_{n_u\times n_u}; A,B\right)
\=\sum_j J\left( I,  P^u_j; A,B\right),\cr
0\=&J\left( I, J; {\bf 1}_{n_d\times n_d},B\right)
\=\sum_a J\left( I,  J; P^d_a, B\right),\cr
0\=&J\left( I, J; A,{\bf 1}_{n_d\times n_d} \right)
\=\sum_b J\left( I,  J; A, P^d_b\right).
\end{aligned} \eeq

\section{A first look}

\subsection{Maximally degenerated Case $G=U(n_u)\times U(n_d)$}

This case has no Jarlskog invariants. The CKM matrix has $n_un_d$ imaginary 
numbers, while the effective symmetry action has dimension 
\beq \underbrace{\dim s(n_u)}_{=\frac{1}{2} n_u(n_u+1)} 
+\underbrace{\dim s(n_d)}_{=\frac{1}{2}n_d(n_d+1)} -\underbrace{1}_{\text{Schur}}
\= n_u n_d +\underbrace{\frac{(n_u-n_d)^2}{2} +\frac{n_u+n_d}{2} -1}_{\geq 0}
,\eeq 
which is always bigger. In fact, singular value decomposition (SVD) 
\beq V\=U_u\begin{pmatrix} \fbox{$\geq 0$} & \cr 
&\fbox{$\geq 0$} & \cr
&&\ddots\cr
&&&\fbox{$\geq 0$} && \cr
\end{pmatrix}_{n_u\times n_d} U^{\dagger}_d\eeq
shows that $V$ is equivalent to a real matrix.

\subsubsection{Unitary case}

If $V$ is unitary it is enough if either the up-masses or the down-masses are
totally degenerate.

\subsection{Counterexample: $2n\times 2n$ matrix with $G=U(n)^2\times U(n)^2$}

The $2n\times 2n$ CKM matrix has $4n^2$ imaginary numbers minus 1 Jarlskog
invariant, while the effective symmetry action has dimension 
\beq 4 \underbrace{\dim s(n)}_{=\frac{1}{2} n(n+1)} 
-\underbrace{1}_{\text{Schur}}.\eeq
Note that the dimension of the symmetry action becomes too small to render a
generic CKM matrix real if $n\geq 2$. 

\subsection{Counterexample: $3n\times 3n$ unitary matrix with 
$G=U(n)^3\times U(n)^3$}

A $3n\times 3n$ unitary CKM matrix has heuristically 
$\dim s(3n)=\frac{1}{2} 3n(3n+1)$ imaginary DOF minus 1 Jarlskog invariant, 
while the effective symmetry action has dimension 
\beq 6 \underbrace{\dim s(n)}_{=\frac{1}{2} n(n+1)} 
-\underbrace{1}_{\text{Schur}}.\eeq
Note that the dimension of the symmetry action becomes too small to render a 
generic unitary CKM matrix real if $n\geq 2$.

\subsection{Example: $G_u=U(n_u)$ and $G_d=U(1)^2$}

\begin{theorem}
If $G_u=U(n_u)$ and $G_d=U(1)^2$, then the $n_u\times 2$ CKM matrix $V$ is 
equivalent to a real matrix.
\label{n2maxnondeg}
\end{theorem}

\noi
{\sc Proof}: Write the $n_u\times 2$ CKM-matrix as
\beq V\=\begin{pmatrix} a & b \cr 
\vec{c} & \vec{d} \end{pmatrix}, \eeq
where $a$ and $b$ constitute the first row, and 
$\vec{c}$ and $\vec{d}$ are column $(n_u\!-\!1)$-vectors.  

\noi
Proceed as follows:
\begin{enumerate}
\item 
Use a $U(n_u)$-transformation to make the column vector 
$\begin{pmatrix} a \cr \vec{c}\end{pmatrix}$ on the form 
$\begin{pmatrix} |a| \cr \vec{0} \end{pmatrix}$ (with possibly a different $a$).
\item 
Use a $U(1)$-rotation to make $b$ real.
\item 
Use a $U(n_u)$-transformation of the form 
$\begin{pmatrix} 1 & 0 \cr 0 & *\end{pmatrix}$
to make $\vec{d}$ real.
\end{enumerate}
\proofbox 

\section{Non-degenerate Case $G=U(1)^{n_u}\times U(1)^{n_d}$}

Introduce polar coordinates 
\beq V^i{}_a\=r^i{}_a\exp(i\theta^i{}_a ) \eeq 
for the matrix elements of the CKM matrix.

\noi
In the non-degenerate case, the Jarlskog invariant reads
\beq \begin{aligned}   J(i,j;a,b)
\=&\Im[V^i{}_a \~(V^j{}_a)^{\ast} \~V^j{}_b \~(V^i{}_b)^{\ast}] \cr
\=&r^i{}_a \~r^j{}_a \~r^j{}_b \~r^i{}_b 
\sin\left( \theta^i{}_a -\theta^j{}_a +\theta^j{}_b -\theta^i{}_b\right).  
\end{aligned}  \eeq 

\begin{definition} Given a CKM matrix $V$ in the non-degenerate case, an 
{\bf echelon cross} is a row and column with non-zero entries only, \cf 
Table~\ref{table11}.
\end{definition}

\begin{table}[ht]
\caption{Example of an echelon cross (marked in red) for a $8\times 7$ CKM 
matrix with only $U(1)$ factors (corresponding to no mass-degeneracies).}
\label{table11}
\begin{center}
\begin{tabular}{|c||c|c|c|c|c|c|c|}  \hline
&\!\!\!\!$U\!(\!1\!)$\!\!\!\!
&\!\!\!\!$U\!(\!1\!)$\!\!\!\!
&\!\!\!\!$U\!(\!1\!)$\!\!\!\!
&\!\!\!\!$U\!(\!1\!)$\!\!\!\!
&\!\!\!\!$U\!(\!1\!)$\!\!\!\!
&\!\!\!\!$U\!(\!1\!)$\!\!\!\!
&\!\!\!\!$U\!(\!1\!)$\!\!\!\!
\cr\hline \hline
$U(1)$ &&&$\red{*}$&&&& \cr \hline 
$U(1)$ &&&$\red{*}$&&&& \cr \hline
$U(1)$ &&&$\red{*}$&&&& \cr \hline  
$U(1)$&$\red{*}$&$\red{*}$&$\red{*}$&$\red{*}$&$\red{*}$ &$\red{*}$&$\red{*}$ 
\cr \hline 
$U(1)$ &&&$\red{*}$&&&& \cr \hline 
$U(1)$ &&&$\red{*}$&&&& \cr \hline
$U(1)$ &&&$\red{*}$&&&& \cr \hline
$U(1)$ &&&$\red{*}$&&&& \cr \hline      
\end{tabular}
\end{center}
\end{table}

\begin{theorem} \cite{cojwk20}
In the non-degenerate case, if a CKM matrix $V$ with vanishing Jarlskog 
invariants $J=0$ contains an echelon cross, then $V$ is equivalent to a real 
matrix.
\label{thmnondeg}
\end{theorem}

\noi
{\sc Proof}: Say that the echelon cross has row number $i$ and column number 
$a$. Use residual $U(1)^{n_u}\times U(1)^{n_d}$ symmetry to make the echelon 
cross real, \ie the corresponding $\theta$-angles $\in\pi\Z$. (Start 
by making the intersection element $(i,a)$ of the cross real. Next make the 
$n_u+n_d-2$ elements in the 4 arms of the cross real.) Finally consider an 
arbitrary element $(j,b)$ outside the cross. The Jarlskog invariant 
$J(i,j;a,b)=0$ is zero. This implies that 
\beq r^j{}_b\=0 \~\~\vee\~\~ \theta^j{}_b\~\in\~\pi\Z, \eeq
\ie the CKM matrix element $V^j{}_b\in\R$ is real.
\proofbox

\subsection{Non-degenerate case $n_u\leq 2\vee n_d\leq 2$}

Interestingly, we don't need an echelon cross for the following 
theorem~\ref{thmnondeg2}.

\begin{theorem}
In the non-degenerate case, if a CKM matrix $V$ with vanishing Jarlskog 
invariants $J=0$ has $n_u\leq 2\vee n_d\leq 2$, then $V$ is equivalent to a 
real matrix.
\label{thmnondeg2}
\end{theorem}

\noi
{\sc Proof}:  Consider \eg the case $n_u=2$.
Consider first the largest submatrix of columns that doesn't contain any zeros.
From Theorem \ref{thmnondeg}, we can assume that this submatrix is real. The 
remaining columns contain at least 1 zero, and can hence be made real by a 
corresponding $U(1)^{n_d}$-rotation.
\proofbox 

\subsection{Counterexample: $3\times 3$ matrix with zero diagonal}

A $3\times 3$ CKM-matrix without an echelon cross is of the form
\beq V\=\begin{pmatrix} 0 & A & \gamma \cr
\beta & 0 & B \cr
C & \alpha & 0 \end{pmatrix} \eeq
up to row and/or column permutations. Here 
$A,B,C,\alpha,\beta,\gamma \in \C$.
The Jarlskog invariants vanish $J=0$. It contains 6 complex phases, but the 
effective residual symmetry group 
\beq \frac{U(1)_L^3\times U(1)_R^3}{U(1)}\eeq 
contains only 5 complex phases, so $V$ is generically 
{\em not}\footnote{However, if \eg the sextic Jarlskog invariant
\beq 
J(1,2,3; 3,1,2) \~ =\~ \Im(\alpha\beta\gamma(ABC)^{\ast})
\eeq vanish, then one can get rid of all complex phases.} equivalent to a real
matrix.

\subsubsection{Unitary case}
\label{sec33unideg}

Since $\det(V)\neq 0$, we conclude that
\beq ABC\~\neq\~0\quad\vee\quad \alpha\beta\gamma\~\neq\~0. \eeq
From the fact that column vectors should be orthogonal it then follows that
\beq (A,B,C)\=(0,0,0)\quad\vee\quad (\alpha,\beta,\gamma)\=(0,0,0), \eeq
\ie there are actually only 3 complex phases, which may easily be removed.

\subsection{Counterexample: $4\times 4$ unitary matrix with zero off-diagonal}

Consider a $4\times 4$ unitary CKM-matrix of the form
\beq V\=\begin{pmatrix} 
* & * & * & 0 \cr
* & * & 0 & * \cr
* & 0 & * & * \cr
0 & * & * & * \end{pmatrix}. \eeq
Only 6 of the 36 Jarlskog invariants are not manifestly zero from the on-set:
\beq J(1,2;1,2), J(1,3;1,3), J(2,4;2,4), J(3,4;3,4), J(1,4;2,3), J(2,3;1,4). 
\eeq 
Nevertheless, the remaining 6 are also zero because of linear relations among 
the Jarlskog invariants, \cf subsubsection \ref{seclinrel}. Let us consider a
unitary CKM-matrix of the form $V=e^{iS}$, where 
\beq S\=\begin{pmatrix} 
* & * & * & 0 \cr
* & * & 0 & * \cr
* & 0 & * & * \cr
0 & * & * & * \end{pmatrix} \eeq
is an infinitesimal real symmetric matrix. One may check that $V$ contains 12
infinitesimal imaginary entries, hereof 8 independent. However the effective
residual symmetry group 
\beq \frac{U(1)_L^4\times U(1)_R^4}{U(1)}\eeq 
contains only 7 complex phases, so $V$ is generically  {\em not} equivalent to
a real matrix.

\section{Allowing degeneracy}

\subsection{Example: $G_u=SU(2)$ and $G_d=U(1)^{n_d}$}

\begin{theorem}
Let\footnote{Here we have cut $G_u$ down to an effective subgroup
$SU(2)\subseteq U(2)$.} $G_u=SU(2)$ and $G_d=U(1)^{n_d}$. Then the $2\times n_d$
CKM matrix $V$ is equivalent to a real matrix if $n_d\leq 2$, but generically 
not\footnote{Theorem~\ref{thm2ndeg} holds for all $n_d$ if additionally all 
the sextic Jarlskog invariants of the form 
\beq \begin{aligned}J(12,12,12;1,2,a)
 \~ = \~ &\Im((V^{\dagger}V)_{12} (V^{\dagger}V)_{2a}(V^{\dagger}V)_{a1})\cr
 \~ = \~ &\Im(\vec{c}_1^{\dagger}\vec{c}_2\~ \vec{c}_2^{\dagger}\vec{c}_a\~
\vec{c}_a^{\dagger}\vec{c}_1)\cr
 \~ = \~ &\vec{c}_1^{\dagger}\vec{c}_2 \Im((A_2 A_a +B_2 B_a) 
(A^{\ast}_a A_1 +B^{\ast}_a B_1))\cr
 \~ = \~ &\underbrace{\vec{c}_1^{\dagger}\vec{c}_2}_{\neq 0}
\underbrace{(A_1B_2-A_2B_1)}_{\neq 0} \Im(A^{\ast}_a B_a) \end{aligned} \eeq
vanish.
Here it is implicitly implied that the column 2-vectors have been prepared as 
indicated in the main proof. We conclude that 
$\arg(B_a)=-\arg(A_a)\in\frac{\pi}{2}\Z$, \ie the column 2-vector 
$\vec{c}_a$ can be made real by a $U(1)$-rotation.} if $n_d\geq 3$.
\label{thm2ndeg}
\end{theorem}

\noi
{\sc Proof}: This case has no Jarlskog invariants. Let us write the 
$2\times n_d$ CKM matrix
\beq V \= \begin{pmatrix}
\vec{c}_1 & \vec{c}_2 & \vec{c}_3 & \ldots &\vec{c}_{n_d} 
\end{pmatrix}, \eeq
in terms of  column 2-vectors. Moreover, let us use the notation 
$\vec{c}=\begin{pmatrix}A \cr B\end{pmatrix}$ for an arbitrary column 2-vector. 

\noi
Preparations: 
\begin{itemize}
\item
In the case $n_d\geq 3$, arrange if possible by column permutations, so that 
$\vec{c}_1$ and $\vec{c}_2$ are neither orthogonal nor parallel. 
(The opposite case goes as follows: Then all column 2-vector can be split into
two orthogonal sets of parallel column 2-vectors. After an $SU(2)$ 
transformation, we may assume that each column 2-vector has a zero component, 
\ie they can all be made real by $U(1)$-rotations.)
\item
By column permutations, we may assume that the first column 2-vector 
$\vec{c}_1$ is non-zero. 
\item
After a $SU(2)$ transformation, we may assume that $\vec{c}_1$ is of the form 
$\begin{pmatrix}A \cr 0\end{pmatrix}$. Using a $U(1)$-rotation, it become of 
the form $\begin{pmatrix}|A| \cr 0\end{pmatrix}$, $A\neq 0$.
\item
For the other column 2-vectors $\vec{c}=\begin{pmatrix}A \cr B\end{pmatrix}$, 
we $U(1)$-rotate so that $A$ and $B$ have opposite arguments, \ie $AB\geq 0$.

\end{itemize}

\noi
Let us now study the effect of an $SU(2)$ transformation 
\beq G_u\=SU(2)\=\left.\left\{ 
\begin{pmatrix} x & y \cr -y^{\ast} &x^{\ast}\end{pmatrix}
\in{\rm Mat}_{2\times 2}(\C) \right| x,y\~\in\~\C, 
\~|x|^2+|y|^2\=1 \right\}.\eeq
on a column 2-vector:
\beq \begin{pmatrix} x & y \cr -y^{\ast} &x^{\ast}\end{pmatrix} 
\begin{pmatrix}A \cr B\end{pmatrix}
\=\begin{pmatrix}xA+yB \cr -y^{\ast}A+x^{\ast}B\end{pmatrix}.
\label{su2column} 
\eeq
A necessary and sufficient condition to achieve a real 2-vector by a 
$U(1)$-rotation is that the 2 components on the RHS of \eq{su2column} must 
have the same phase (modulo $\pi$), \ie
\beq \R\~\ni\~(xA+yB)(-y^{\ast}A+x^{\ast}B)^{\ast} 
\=xy(|B|^2-|A|^2) +x^2AB^{\ast} -y^2 BA^{\ast}. \label{realcond01} \eeq
From the first column 2-vector, we conclude that $x$ and $y$ have opposite 
phases (modulo $\pi$). In particular $x^2$ and $y^2$ have from now on opposite
arguments.

\noi
In the following we will implicitly assume that $x\neq 0$. (For the case $x=0$,
one can instead give an argument using $y\neq 0$ in a very similar fashion.)

\noi
Consider now another column 2-vector (different from the first). We may assume
that $AB\neq 0$, because else the necessary and sufficient condition is already
satisfied. The imaginary part of the RHS of \eq{realcond01} becomes
\beq 0\=\Im (RHS)
\=(|x|^2+|y|^2)|AB|\sin\arg (x^2AB^{\ast})\quad \Leftrightarrow \quad
\arg (x^2AB^{\ast})\~\in\~\pi\Z. \eeq
By choosing $\arg(x)$ this is always possible to satisfy for a given second 
column 2-vector, but generically impossible for more column vectors.
\proofbox

\subsection{$2\times 2$ CKM matrix}

\begin{theorem}
A $2\times 2$ CKM matrix where all Jarlskog invariants vanish is equivalent to
a real matrix.
\label{thm22}
\end{theorem}

\noi
{\sc Proof}: 
\begin{itemize}
\item Non-degenerate case: Use theorem~\ref{thmnondeg2}.
\item Degenerate case: There are no Jarlskog invariants.
By symmetry it is enough to consider the degenerate case where 2 up-quarks 
have the same mass. Use theorem~\ref{thm2ndeg}.
\end{itemize}
\proofbox

\subsection{$3\times 3$ unitary CKM matrix}
\label{sec33uni}

This is the standard model case.\footnote{The $3\times 3=9$ non-degenerate
Jarlskog invariants 
\beq J(i,j;a,b) \~ =\~ J \sum_{k,c} \epsilon_{ijk} \epsilon_{abc}\eeq
are alternating versions of a single invariant $J$. The invariant
\cite{j85a,j85b,js88,bn87,j89,f09} 
\beq \Im\det[Y_u,Y_d] \~\propto\~J\Delta(m^u_i)\Delta(m^d_a)\eeq
takes mass-degeneracy into account. Here $Y_u$, $Y_d$ are Yukawa matrices, and 
\beq \Delta(m_i)\~ = ~\prod_{i<j} (m_i\!-\!m_j)\eeq is the Vandermonde 
determinant.}

\begin{theorem}
A $3\times 3$ unitary CKM matrix where all Jarlskog invariants vanish is
equivalent to a real matrix.
\label{thm33}
\end{theorem}

\noi
{\sc Proof}: 
For the non-degenerate case, use theorem~\ref{thmnondeg} if $V$ has an echelon
cross, and subsubsection~\ref{sec33unideg} if it doesn't. Next let's consider
the degenerate cases. Then there are no Jarlskog invariants. For this reason
it is enough to consider the smallest degenerate symmetry group, \eg 
\beq G_u\=U(2)\times U(1)\quad\text{and}\quad G_d\=U(1)^3.\eeq
Let the unitary CKM matrix be
\beq V\= \begin{pmatrix} A & C & E \cr 
B & D & F \cr 
\alpha & \beta & \gamma \end{pmatrix}. \eeq
Proceed as follows:
\begin{enumerate}
\item 
We can assume (after possibly permuting columns) that the column 2-vector 
$\begin{pmatrix} A \cr B\end{pmatrix}\neq\vec{0}$ is non-zero. 
\item 
Use $U(1)_R$-rotations to make the 3rd row $\alpha$, $\beta$ and $\gamma$ real.
\item 
Use a $U(2)_L$-transformation to make the column 2-vector 
$\begin{pmatrix} A \cr B\end{pmatrix}$ on the form 
$\begin{pmatrix} |A| \cr 0 \end{pmatrix}$.
\item 
Use orthogonality of the column vectors to conclude that $C$ and $E$ are real.
\item 
Use orthogonality of the column vectors to conclude that $D$ and $F$ have the
same phase (modulo $\pi$).
\item 
Use a $U(2)_L$-transformation of the form 
$\begin{pmatrix} 1 & 0\cr 0 & * \end{pmatrix}$ to make $D$ and $F$ real.
\end{enumerate}
\proofbox 

\section{Case of at most 2-fold mass-degeneracies, \ie only $U(1)$ and $U(2)$ 
factors}

By permuting rows and columns, we may assume that all $U(2)$-factors are
ordered before the $U(1)$-factors, \ie
\beq G_u\=U(2)^{m_u}\times U(1)^{n_u-2m_u}, \qquad 
m_u\~\in\~\{1,2,\ldots, [\frac{n_u}{2}] \},\eeq 
\beq G_d\=U(2)^{m_d}\times U(1)^{n_d-2m_d}, \qquad 
m_d\~\in\~\{1,2,\ldots, [\frac{n_d}{2}] \}.\eeq 

\begin{table}[ht]
\caption{Example of an echelon cross (marked in red) for a 
$11\times 10$ CKM matrix with only $U(1)$ and $U(2)$ factors.
The echelon partners and echelon copartners are marked in blue,
 while the echelon children are marked in green.}
\label{table12}
\begin{center}
\begin{tabular}{|c||c|c|c|c|c|c|c|c|}  \hline
& $U(2)$ & $U(2)$
&\!\!\!\!$U\!(\!1\!)$\!\!\!\!
&\!\!\!\!$U\!(\!1\!)$\!\!\!\!
&\!\!\!\!$U\!(\!1\!)$\!\!\!\!
&\!\!\!\!$U\!(\!1\!)$\!\!\!\!
&\!\!\!\!$U\!(\!1\!)$\!\!\!\!
&\!\!\!\!$U\!(\!1\!)$\!\!\!\!
\cr\hline \hline
\multirow{2}{*}{$U(2)$} &  & & $\red{*}$ && $\blue{*}$  &&& \cr 
 &  & & $\red{*}$ && $\blue{*}$  &&& \cr \hline 
\multirow{2}{*}{$U(2)$} &  & & $\red{*}$ &&&& $\blue{*}$&   \cr 
 &  & & $\red{*}$ &&&& $\blue{*}$&   \cr \hline
\multirow{2}{*}{$U(2)$} &  & & $\red{*}$ & $\blue{*}$&&&&   \cr 
 &  & & $\red{*}$ & $\blue{*}$ &&&&  \cr \hline 
$U(1)$ &$\red{*\~\~*}$&$\red{*\~\~*}$
&$\red{*}$&$\red{*}$&$\red{*}$ &$\red{*}$&$\red{*}$&$\red{*}$ \cr \hline 
$U(1)$ &&&$\red{*}$&&&&& \cr \hline 
$U(1)$ &&$\blue{*\~\~*}$
&$\red{*}$&$\green{*}$&$\green{*}$&&$\green{*}$& \cr \hline 
$U(1)$ &&&$\red{*}$&&&&& \cr \hline 
$U(1)$ &$\blue{*\~\~*}$&
&$\red{*}$&$\green{*}$&$\green{*}$&&$\green{*}$& \cr \hline  
\end{tabular}
\end{center}
\end{table}

\begin{definition}  ~\\
\begin{itemize}
\item
A singlet-singlet $1\times 1$ matrix element is called {\bf echelon} if it is
non-zero.
\item
A doublet-singlet $2\times 1$ submatrix $\vec{v}$ is called {\bf echelon} if it
has an echelon partner. An {\bf echelon partner}\footnote{Echelon partners are 
often marked in blue in this article, \cf Table~\ref{table12}. The reader may
wonder why we don't allow an echelon partner to be a doublet-doublet $2\times 2$
submatrix. The short answer is that it turns out to not be practical/useful.
See also the analysis in subsection~\ref{secnexttomax}.}
is another doublet-singlet $2\times 1$ submatrix $\vec{w}$ within the same
doublet-row such that $\vec{v}$ and $\vec{w}$ are neither parallel nor
perpendicular, \ie $\det[\vec{v},\vec{w}]\neq 0$ and
$\vec{v}^{\dagger}\vec{w}\neq 0$.
\item
A singlet-doublet $1\times 2$ submatrix $\vec{v}^T$ is called {\bf echelon} if
it has an echelon co-partner. An {\bf echelon co-partner} is another
singlet-doublet $1\times 2$ submatrix $\vec{w}^T$ within the same
doublet-column such that $\vec{v}$ and $\vec{w}$ are neither parallel nor
perpendicular. 
\item
A doublet-doublet $2\times 2$ submatrix is never {\bf echelon}. 
\end{itemize}
\end{definition}

\begin{definition} For each pair of echelon (partner, co-partner), the
singlet-singlet $1\times 1$ matrix element in the same column as the partner,
and in the same row as the co-partner, is called an {\bf echelon child}. 
\end{definition}

\begin{definition} Given a CKM matrix $V$ of the singlet-deplete type, an
{\bf echelon cross} is a singlet row and a singlet column\footnote{Echelon
crosses are often marked in red in this article, \cf Table~\ref{table12}. 
The reader may wonder why we don't define an echelon cross built from a 
double-row and/or a double-column of doublet type? The short answer is that 
it turns out to not be practical/useful. Try!} with echelon block entries 
only, and such that all echelon children are non-zero.
\end{definition}

\begin{theorem}
If a CKM matrix $V$ of the singlet-deplete type with vanishing Jarlskog
invariants $J=0$ contains an echelon cross, then $V$ is equivalent to a real
matrix.
\label{thm12}
\end{theorem}

\noi
{\sc Sketched proof}:
Use repeatedly the following examples~\ref{ex1}-\ref{ex3}.
\proofbox

\subsection{Example: $G_u=U(2)\times U(1)$ and $G_d=U(1)^2$}
\label{ex1}

Assume that the 3rd row and 1st column constitute an echelon cross.
\beq V\=\begin{pmatrix}
\red{\vec{a}} & \blue{\vec{b}} \cr\red{c} & \red{d}
\end{pmatrix}. \eeq
The Jarlskog invariant is
\beq 0\=J(12,3;1,2)\=\Im (\vec{b}^{\dagger}\vec{a}c^{\ast}d).\eeq
We can assume that the column 2-vectors $\vec{a}$ and $\vec{b}$ are real 
because of theorem~\ref{thm22}.
From the Jarlskog invariant, we see that $c$ and $d$ must have the same phase
(modulo $\pi$). (Here we have used that $\vec{a}$ and $\vec{b}$ are not
perpendicular.) We can hence apply a $U(1)_L$-rotation on the 3rd row to make
it real.
\proofbox

\subsection{Example: $G_u=U(2)\times U(1)$ and $G_d=U(1)^3$}
\label{ex2}

Assume that the 3rd row and the 1st column is an echelon cross.
\beq  V\=\begin{pmatrix}
\red{A} & \blue{C} & E \cr\red{B} & \blue{D} & F \cr
\red{\alpha} & \red{\beta} & \red{\gamma} 
\end{pmatrix}. \eeq
We can assume that the 2 first column 3-vectors are real because of 
example~\ref{ex1}. By applying a $U(1)_R$-rotation we can assume that $E$ and
$F$ have opposite arguments. The pertinent Jarlskog invariants are
\beq \begin{aligned}
0\=&J(12,3;1,3)\~\propto\~\Im ((AE +BF)\gamma^{\ast}),\cr
0\=&J(12,3;2,3)\~\propto\~\Im ((CE + DF)\gamma^{\ast}).
\end{aligned}\eeq
We conclude that $AE + BF$, $CE + DF$ and $\gamma$ must have the same phase
(modulo $\pi$). In particular,
\beq  \R\~\ni\~(AE + BF)(CE + DF)^{\ast}, \eeq
or equivalently,
\beq 0\=\Im (ADEF^{\ast} +BCFE^{\ast})
\=\underbrace{(AD-BC)}_{\neq 0}|EF| \sin (2\arg(E)). \eeq
There are 2 cases:
\begin{itemize}
\item $E=0 \vee F=0$.
\item $\arg{E}\in\frac{\pi}{2}\Z$.
\end{itemize}

\noi
In both cases we can make $E$ and $F$ real by applying a $U(1)_R$-rotation. 

\noi
There are 2 cases:
\begin{itemize}
\item $(E,F) \neq (0,0)$. Both inner products $AE + BF$ and $CE + DF$ cannot
be zero. Hence $\Im (\gamma)=0$.

\item $(E,F) = (0,0)$. Make $\gamma$ real by applying a $U(1)_R$-rotation.
\end{itemize}
\proofbox

\subsection{Example: $G_u=U(2)\times U(1)^2$ and $G_d=U(2)\times U(1)^2$}
\label{ex3}

Assume that the 3rd row and 3rd column is an echelon cross.
\beq V\=\begin{pmatrix}
A & \red{\vec{c}} & \blue{\vec{d}} \cr 
\red{\vec{b}^T} & \red{\beta} & \red{\gamma} \cr
\blue{\vec{a}^T} & \red{\alpha} & \green{*} 
\end{pmatrix}. \eeq
We can assume that the column 2-vectors $\vec{a}$, $\vec{b}$, $\vec{c}$ and
$\vec{d}$ are real because of theorem~\ref{thm22}.

\noi
We can assume that phases of $\alpha$, $\beta$ and $\gamma$ are the same
(modulo $\pi$) because of example~\ref{ex1}. By applying the same 2
$U(1)_R$-rotations (and an opposite central $U(1)_L$-rotation inside $U(2)_L$),
we can make $\alpha$, $\beta$ and $\gamma$ real (without disturbing $\vec{c}$
and $\vec{d}$). Now the echelon cross is real.

\noi
From the non-degenerate theory, we can make the echelon child $*$ real.

\noi
The 4 pertinent Jarlskog invariants are
\beq \begin{aligned}
0\=&J(12,3;12,3)\~\propto\~\vec{c}^T\Im (A)\vec{b},\cr
0\=&J(12,4;12,3)\~\propto\~\vec{c}^T\Im (A)\vec{a},\cr
0\=&J(12,3;12,4)\~\propto\~\vec{d}^T\Im (A)\vec{b},\cr
0\=&J(12,4;12,4)\~\propto\~\vec{d}^T\Im (A)\vec{a},
\end{aligned}\eeq
which are 4 independent linear equations for $\Im (A)$. (The 4 conditions can
be solved more easily if we use $U(2)$ transformation to make the 2nd
components of $\vec{a}$ and $\vec{b}$ equal to zero.) We conclude that
$\Im (A)=0$.
\proofbox

\section{Supplementary Material}

\subsection{Higher Echelon Subblocks?}

The strategy so far has been to divide the question [of whether a CKM matrix 
is equivalent to a real matrix] into (i) a generic case where an echelon 
cross guarantees this, and (ii) a special case of Lebesgue-measure zero
where further analysis is needed.

\noi
It is therefore natural to try to generalize echelon entries to $n$-fold 
degeneracy along the following lines.

\begin{definition} 
A $n$-plet-singlet $n\times 1$ submatrix $\vec{v}$ is called {\bf echelon}
if it has $n\!-\!1$ echelon partners.
An {\bf echelon partner}
is another $n$-plet-singlet $n\times 1$ submatrix $\vec{w}$ within the same
$n$-plet-row that satisfies the following conditions: The $n$
$n$-plet-singlets are linearly independent, but pairwise not perpendicular.
\end{definition}

\noi
However, this will not be useful as we shall see below in the case $n=3$.

\subsection{Example: $G_u=SU(3)$ and $G_d=U(1)^3$}

Let's write the $3\times 3$ CKM matrix as
\beq V\=\begin{pmatrix}
\vec{a} & \vec{b} & \vec{c} 
\end{pmatrix}, \eeq
where $\vec{a}$, $\vec{b}$, and $\vec{c}$ are linearly independent column 
3-vectors that are pairwise not perpendicular.

\noi
We can wlog.\ assume that $\vec{a}$ and $\vec{b}$ are real, \cf 
theorem~\ref{n2maxnondeg}.

\noi
Let us parametrize the $3\times 3$ special unitary matrices as\footnote{
Here we have cut $G_u$ down to an effective subgroup $SU(3)\subseteq U(3)$.}
\beq U\=e^{iS} O\~\in \~SU(3), \eeq
where $O\in SO(3)$ is a $3\times 3$ orthogonal matrix and $S$ is a traceless 
real symmetric $3\times 3$ matrix, \cf subsection~\ref{secunitarymatrixdecomp}.
Since orthogonal matrices does not change the fact that $\vec{a}$ and 
$\vec{b}$ are real, we can ignore them in what follows.

\subsubsection{Infinitesimal Analysis}

At this point, we assume that $\Im (\vec{c})$ is infinitesimal. Let us imagine
that we successfully perform an infinitesimal symmetry transformation
$U=e^{iS}$,
where $S$ is an infinitesimal real symmetric $3\times 3$ matrices, such that
the CKM matrix becomes real after pertinent infinitesimal $U(1)^3$-rotations.
This implies that 
\beq \begin{aligned}
\vec{a}^{\prime}\~:=\~e^{iS}\vec{a}\qquad\Rightarrow&\qquad
\Im(\vec{a}^{\prime})\=S\vec{a} \~\parallel\~ \vec{a},\cr
\vec{b}^{\prime}\~:=\~e^{iS}\vec{b}\qquad\Rightarrow&\qquad
\Im(\vec{b}^{\prime})\=S\vec{b} \~\parallel\~ \vec{b} ,\cr
\vec{c}^{\prime}\~:=\~e^{iS}\vec{c}\qquad\Rightarrow&\qquad
\Im(\vec{c}^{\prime})\=S \Re(\vec{c}) + \Im(\vec{c})
 \~\parallel\~ \Re(\vec{c}).
\label{propto3111}  
\end{aligned}\eeq
The 3 eqs.~(\ref{propto3111}) contain 3 proportionality constants 
$(\lambda_a,\lambda_b,\lambda_c)$. This means that we have $3+3+3=9$ real 
equations, but only $5+3=8$ real unknowns $(S,\lambda_a,\lambda_b,\lambda_c)$. 
This does {\em not}\footnote{However the DOF matches if we take into account 
that there is precisely 1 independent higher Jarlskog invariant, \eg
\beq\begin{aligned}J(123,123,123;1,2,3)
\~ = \~ &\Im((V^{\dagger}V)_{12} (V^{\dagger}V)_{23}(V^{\dagger}V)_{31})\cr
\~ = \~ &\Im(\vec{a}^{\dagger}\vec{b}\~\vec{b}^{\dagger}\vec{c}\~
\vec{c}{}^{\dagger}\vec{a})
\~ = \~ \vec{a}^{\dagger}\vec{b}\~
\Im(\vec{b}^{\dagger}\vec{c}\~\vec{c}{}^{\dagger}\vec{a}) \cr
\~ = \~ & \vec{a}^{\dagger}\vec{b}\left(
\vec{b}^{\dagger}\Im(\vec{c})\~\vec{a}^{\dagger}\Re(\vec{c})
-\vec{b}^{\dagger}\Re(\vec{c})\~\vec{a}^{\dagger}\Im(\vec{c})\right)
.\end{aligned} \eeq} have solutions in general. 
(Since $\vec{a}$ and $\vec{b}$ are non-perpendicular eigenvectors to the
traceless real symmetric matrix $S$, it follows that their eigenvalues
$\lambda_a=\lambda_b$ are equal. Nevertheless, this fact does {\em not} mean 
that we are short of 2 DOF rather than 1 DOF.)

\subsection{Next-to-maximally degenerate case}
\label{secnexttomax}

\noi
The symmetry group is assumed to be
\beq G_u\=U(n_u\!-\!1)\times U(1)
\quad\text{and}\quad 
G_d\=U(n_d\!-\!1)\times U(1).\eeq
We may wlog.\ assume that $2\leq n_u\leq n_d$. 
Let's write the $n_u\times n_d$ CKM matrix as
\beq V\=\begin{pmatrix}
A & \vec{b} \cr 
\vec{c}^T & d
\end{pmatrix}, \eeq
where we assume that  $A$ is a $(n_u\!-\!1)\times (n_d\!-\!1)$ matrix with a 
right inverse, $\vec{b}$ is a non-zero column $(n_u\!-\!1)$-vector, $\vec{c}$ 
is a non-zero column $(n_d\!-\!1)$-vector, and $d\in\C\backslash\{0\}$ is a 
non-zero number. (In other words, we assume for simplicity that the 4 
subblocks have maximal rank.)

\noi
First use SVD to make $A$ and $d$ non-negative and diagonal. By left and right
diagonal $U(1)$-rotations, we can make $\vec{c}$ real. In particular,
\beq A, \vec{c},\text{ and }d\text{ are real}.\eeq 
There is 1 Jarlskog invariant:
\beq J(12\ldots ,n_u;12\ldots,n_d)
\=-\Im (\vec{b}^{T}A^{\ast} \vec{c} d^{\ast})
\~\propto\~\Im (\vec{b})^{T}A\vec{c}. \eeq
If $n_u\leq 2$ and if $A\vec{c}\neq \vec{0}$, then we can conclude that 
$\Im(\vec{b})=\vec{0}$, so that $V$ is real, and we're done. Let us therefore 
assume that $n_u\geq 3$.

\noi
Let us parametrize the unitary matrices as
\beq U\=e^{iS}O, \eeq
where $O$ is an orthogonal matrix and $S$ is a real symmetric matrix, \cf 
subsection~\ref{secunitarymatrixdecomp}. Since orthogonal matrices does not 
change the fact that $A$, $\vec{c}$ and $d$ are real,
we can ignore them in what follows. We can also ignore the last $U(1)$-factor
in both groups $G_u$ and $G_d$.

\subsubsection{Infinitesimal Analysis}

At this point, we assume that $\Im (\vec{b})$ is infinitesimal. Let us imagine
that we successfully perform an infinitesimal symmetry transformation
\beq U_u\= \begin{pmatrix}
e^{iS_u} & 0 \cr 
0 & 1 \end{pmatrix}, \qquad U_d\= \begin{pmatrix}
e^{iS_d} & 0 \cr 
0 & 1 \end{pmatrix}, \eeq
where $S_u$ and $S_d$ are infinitesimal real symmetric matrices, such that 
$V^{\prime}=U_u VU_d^{\dagger}$ is real.
\beq \begin{aligned}
0\=&\Im(A^{\prime})\=S_u A-AS_d
\quad\Leftrightarrow\quad S_u\=AS_dA^{-1},\cr
0\=&\Im(\vec{b}^{\prime})
\=S_u Re(\vec{b}) +\Im(\vec{b})\=AS_dA^{-1} Re(\vec{b}) +\Im(\vec{b}),\cr
0\=&\Im(\vec{c}^{\prime})\=S_d \vec{c}.  
\end{aligned}\eeq
If it happens that $A^{-1} Re(\vec{b})\parallel\vec{c}\neq\vec{0}$ and if 
$\Re(\vec{b})\perp\Im(\vec{b})\neq\vec{0}$ then $S_d$ does not exist. We 
conclude that a vanishing Jarlskog invariant does {\em not} guarantee that 
the CKM matrix $V$ is equivalent to a real matrix if $n_u\geq 3$.

\subsection{The case $G=U(2)^2\times U(2)^2$ with invertible $2\times 2$ 
sub-blocks}
\label{sec2222}

Let a $4\times 4$ CKM matrix be of the form
\beq V\=\begin{pmatrix}
A & B \cr 
C & D
\end{pmatrix}, \eeq
where $A,B,C,D$ are invertible $2\times 2$ matrices. There is 1 Jarlskog 
invariant
\beq J(12,34;12,34)\=\Im (\tr(AC^{\dagger}DB^{\dagger})). \eeq
First use SVD to make $B$ and $C$ positive and diagonal. In particular, 
\beq B\text{ and }C\text{ are real}.\eeq
Let us parametrize the $2\times 2$ unitary matrices as
\beq U\=e^{iS} O\~\in \~U(2), \eeq
where $O\in O(2)$ is a $2\times 2$ orthogonal matrix and $S$ is a real 
symmetric $2\times 2$ matrix, \cf subsection~\ref{secunitarymatrixdecomp}. 
Since orthogonal matrices does not change the fact that $B$ and $C$ are real, 
we can ignore them in what follows.

\subsubsection{Infinitesimal Analysis}

At this point, we assume that $\Im (A)$ and $\Im (D)$ are infinitesimal.
Let us imagine that we successfully perform an infinitesimal symmetry 
transformation
\beq U_u\= \begin{pmatrix}
e^{iS_u} & 0 \cr 
0 & e^{iR_u} \end{pmatrix}, \qquad 
U_d\= \begin{pmatrix}
e^{iS_d} & 0 \cr 
0 & e^{iR_d} \end{pmatrix}, \eeq
where $S_u$, $R_u$, $S_d$ and $R_d$ are infinitesimal real symmetric 
$2\times 2$ matrices, such that $V^{\prime}=U_u VU_d^{\dagger}$ is real.
\beq \begin{aligned}
0\=&\Im(A^{\prime})\=S_u\Re(A)-\Re(A)S_d+\Im(A),\cr
0\=&\Im(B^{\prime})\=S_uB-BR_d
\quad\Leftrightarrow\quad 
R_d\=B^{-1}S_uB,\cr
0\=&\Im(C^{\prime})\=R_uC-CS_d
\quad\Leftrightarrow\quad 
S_d\=C^{-1}R_uC,\cr
0\=&\Im(D^{\prime})\=R_u\Re(D)-\Re(D)R_d+\Im(D).  
\end{aligned}\eeq
Eliminating $S_d$ and $R_d$, we get
\beq \begin{aligned}
-\Im(A)\=&S_u\Re(A)-\Re(A)C^{-1}R_uC,\cr
-\Im(D)\=&R_u\Re(D)-\Re(D)B^{-1}S_uB.  
\end{aligned}\eeq
Eliminating $K_u$ leads to
\beq
\Im(A)+S_u\Re(A)\=\Re(A)C^{-1} \left[\Re(D)B^{-1}S_uB-\Im(D)\right]\Re(D)^{-1}C.
\eeq
Multiplying from right with $C^{-1}\Re(D)B^{-1}$ leads to
\beq
\left[\Im(A)+S_u\Re(A)\right]C^{-1}\Re(D)B^{-1}
\=\Re(A)C^{-1} \left[\Re(D)B^{-1}S_u-\Im(D)B^{-1}\right].
\eeq
Taking trace yields the following consistency condition
\beq \tr (\Im(A)C^{-1}\Re(D)B^{-1}) + \tr (\Re(A)C^{-1}\Im(D)B^{-1})\=0. \eeq
This is generically different from the condition that the infinitesimal 
Jarlskog invariant 
\beq 0\=J(12,34;12,34)
\=\tr (\Im(A)C^T\Re(D)B^T) + \tr (\Re(A)C^T\Im(D)B^T) \eeq
vanishes. We conclude that a vanishing Jarlskog invariant does {\em not} 
guarantee that the CKM matrix $V$ is equivalent to a real matrix.

\subsubsection{Discussion}

The above consistency condition suggests that the relevant invariant uses 
inverse matrix operations rather than Hermitian adjoint:
\beq \widetilde{J}(12,34;12,34)\=\Im (\tr(AC^{-1}DB^{-1})). \eeq
However, this would not work for non-invertible blocks. For a similar reason, 
we must restrict the symmetry group to unitary groups rather than general 
linear groups.

\vspace{0.8cm}

\noi
{\sc Acknowledgement:}~The work of K.B.\ is supported by the Czech Science 
Foundation (GACR) under the grant no.\ GA20-04800S for Integrable Deformations.

\end{document}